\documentstyle[epsf]{laa}
\begin{document}
\thesaurus{08.02.1, 08.05.3, 08.09.2: HD\,44179, 08.16.4, 09.04.1}
\title{Variability and nature of the binary in the Red Rectangle Nebula
\thanks{based on observations collected at the European Southern Observatory
and at La Palma Observatory}}
\author{ C. Waelkens\inst{1}, H. Van Winckel\inst{1}, L.B.F.M. Waters\inst{2,3}
and E.J. Bakker\inst{4,5}}
\offprints{C. Waelkens, Leuven}
\institute{
Instituut voor Sterrenkunde, K.U. Leuven, Celestijnenlaan 200B,
3001 Heverlee, Belgium \and
Astronomisch Instituut, Universiteit van Amsterdam, Kruislaan 403,
1098 SJ Amsterdam, The Netherlands \and
SRON Groningen, P.O. Box 800, 9700 AV Groningen, The Netherlands \and
SRON Utrecht, Sorbonnelaan 2, 3584 CA Utrecht, The Netherlands \and
Astronomical Institute, University of Utrecht, P.O. Box 80000,
3508 TA Utrecht, The Netherlands}

\date{received\dots, accepted\dots}

\maketitle
\markboth{C. Waelkens et al.:
Variability and nature of the binary in the Red Rectangle Nebula}
{C. Waelkens et al.:
the binary in the Red Rectangle}

\begin{abstract}

We present new observations of the central binary inside the Red Rectangle
nebula.  The detection of zinc in the optical spectrum confirms that the
peculiar photospheric abundances are due to accretion of circumstellar gas.
Grey brightness variations with the orbital period are observed.  They are
interpreted as being due to the  variation of the scattering angle with orbital
phase.  The small orbital separation of the system is not compatible with
previous normal evolution of the primary on the AGB.  We point out the 
similarity of the orbital history of this and other similar systems with those 
of some close Barium stars and suggest that the nonzero eccentricity of the 
orbit is the result of tidal interaction with the circumbinary disk.


\end{abstract}

\section{Introduction}
The Red Rectangle nebula that surrounds the star HD\,44179 (Cohen et al.
1975) is famous for the molecular and dusty emission it displays in the red
and infrared parts of the spectrum.  The central star is a peculiar
A-supergiant.  A major puzzle is that the infrared-to-optical luminosity of
the central object in the Red Rectangle is very high (about 33, Leinert and
Haas 1989), despite the fact that the extinction of HD\,44179 is not larger
than \(E(B-V)\,=\,0.4\).  This led Rowan-Robinson and Harris (1983) to invoke
the presence of an embedded M-giant companion; Leinert and Haas went even
further, and argued that HD\,44179 is a foreground object.  The solution of
this puzzle is contained in the observations by Roddier et al. (1995), who
showed that the optical flux observed from HD~44179 is entirely 
scattered light from 
two lobes located above and below a dusty disk, the star itself being hidden 
by the disk.  There is then no need to invoke another source than HD~44179 
to power the luminosity of the nebula.  

Waelkens et al. (1992) have shown that HD\,44179 is severely iron-deficient, 
with \( [Fe/H]\,=\,-3.3\), that also other metals such as Mg, Si, and Ca are
severely underabundant, but that the CNO and S abundances of this star are
nearly solar.  This star is then clearly of the same nature as the other
extremely iron-poor supergiants HR~4049, HD~52961 and BD +39$^{\circ}$4926
(Lambert et al. 1988; Waelkens et al. 1991a; Kodaira 1973; Bond 1991), which
show the same most peculiar abundance pattern, and two of which (HR~4049 and
HD~52961) are also surrounded by circumstellar dust.  The location of these 
other low-gravity stars rather far from the galactic plane strongly suggests 
that they are not massive supergiants, but evolved low-mass stars, and thus 
that the central star of the Red Rectangle also is an evolved low-mass star.
Moreover, the dust features show clearly that the nebula is carbon-rich, which
argues that the star has undergone the third dredge-up
typical for late AGB stars.  Still, the large amount of circumstellar matter
and the low galactic latitude \(b\,=\,-12^{\circ}\) may indicate that HD\,44179
is somewhat more massive than the other stars of the group.  The carbon
richness of the circumstellar environment, as well as the high luminosity 
combined with the high or intermediate galactic latitudes, suggest that
these stars are low- or intermediate-mass objects in a post-AGB stage of evolution.

While there is no ground any more to consider HD~44179 as a component of a
{\it wide} binary, Van Winckel et al. (1995) have shown that HD\,44179 is a
{\it spectroscopic} binary with an orbital period of about 300 days.  Also the
other mentioned extremely iron-poor post-AGB stars are binaries with periods 
of the order of one year.  

\begin{table*}
\caption{Photospheric abundances of the extremely iron-poor post-AGB stars
(the zinc abundance for HD~44179 is from the present study, the other values
are taken from Van Winckel 1995)}
\begin{tabular}{l|lrrrrrrrr}
\hline
Object & Iras & [Fe/H] & [C/H] & [N/H] & [O/H] & [S/H] & [Mg/H] & [Si/H] & [Zn/H] \\
\hline
HR 4049 & 10158$-$2844 &  -4.8 & -0.2 & 0.0 & -0.5 & -0.4 & ? & ? & ? \\
HD 52961 & 07008$+$1050 & -4.8 & -0.4 & -0.4 & -0.6 & -1.0 & ? & ? & -1.5 \\
BD +39$^{\circ}$4926 & & -3.3 & -0.3 & -0.4 & -0.1 & 0.1 & -1.5 & -1.9 & ? \\
HD 44179 & 06176$-$1036 & -3.3 & 0.0 & 0.0 & -0.4 & -0.3 & -2.1 & -1.8 & -0.6 \\
\hline
\end{tabular}
\end{table*}

In this paper we discuss new observations of HD~44179 that were triggered by
the similarity of this object with the other peculiar binaries.  In Section~2
we report on the determination of the zinc abundance in HD~44179, which
confirms that the photospheric peculiarity is due to accretion of circumstellar
gas.  In Section~3 we discuss the photometric variability of HD~44179; for
HR~4049 the brightness and colours vary with orbital phase, as a result of
variable circumstellar absorption; in HD~44179 it appears more likely that the
observed variability is due to periodically variable scattering.  In Section~4
we discuss the constraints the close-binary nature of HD~44179 imposes on the
previous and future evolution of this object.

\section{The zinc abundance of HD~44179}

In Table\,1 we summarize the photospheric composition of the four objects
that are known to belong to the group of extremely iron-poor post-AGB stars.
These stars are characterized by very low abundances of refractory elements
such as Fe, Mg and Si, and about normal abundances of CNO and S.  Following
a suggestion by Venn and Lambert (1990), Bond (1991) first suggested that
the low iron abundances are not primordial, but are due to fractionation
onto dust.  Indeed, the abundance pattern of these stars follows that of the
interstellar gas rather closely.

Convincing evidence for this scenario came from the detection of a rather
solar Zn abundance in HD\,52961 by Van Winckel et al. (1992): in the
photosphere of this star, there is more zinc than iron in absolute numbers!
The unusual zinc-to-iron ratio cannot be understood in terms of
nucleosynthetic processes, but must be due to the different {\it chemical}
characteristics of both elements, zinc having a much higher condensation
temperature than iron (Bond 1992).

From the study of a spectrum of HD~44179 obtained with the Utrecht Echelle
Spectrograph at the William Herschel Telescope at La Palma, we can confirm
that zinc follows CNO and S also in this star.  In Figure~1 we show a spectrum
with the 4810 $\AA$\ zinc line for HD~44179.  The zinc abundance derived from 
this line and the 4722 $\AA$\ line is [Zn/H] = -0.6, while [Fe/H] = -3.3.  
Also in the photosphere of HD\,44179 the amount of zinc is near that of iron 
in absolute numbers.  This result further underscores that the central star of 
the Red Rectangle belongs to the same group of objects as HR\,4049, HD\,52961 
and BD+39$^{\circ}$4926.  The fact that all four such objects known occur in 
binaries with similar periods then strengthens the suggestion by Waters et al. 
(1992) that the chemical separation process occurs in a 
stationary {\it circumbinary} disk.

\begin{figure}
\mbox{\epsfxsize=3.2in\epsfysize=2.2in\epsfbox[55 40 550
780]{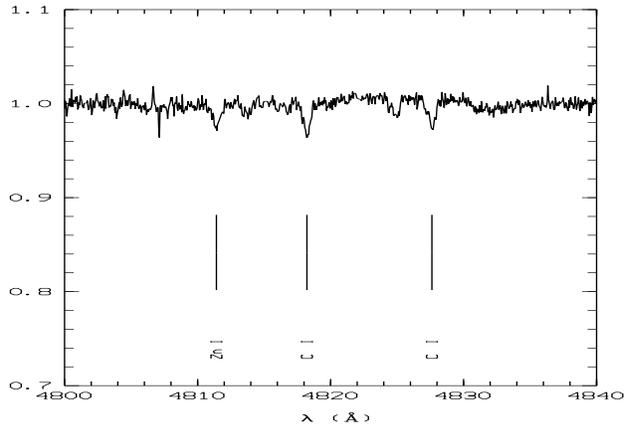}}
\caption[]{The observation of a zinc line in the spectrum of HD\,44179}
\end{figure}

\section{Photometric variability}

Seen at a high inclination, a circumbinary disk can cause variable
circumstellar extinction, because the amount of dust along the line of sight
varies with the orbital motion of the star.  Such an effect has indeed been
observed for HR\,4049 (Waelkens et al. 1991b).  We have therefore obtained
68 photometric measurements in the Geneva photometric system, with the Geneva
photometer attached to the 0.70 Swiss Telescope at La Silla Observatory in
Chile, between 1992 and 1996, covering now six orbital cycles. In order to
improve on the orbital elements, we have also obtained new radial-velocity 
measurements with the CES spectograph fed by the CAT telescope at ESO; the 
data now cover more than five cycles.  In Figure~2 we fold the observed visual 
magnitudes and [U-B] colors with the orbital phase.  The orbital period of 
318\,$\pm$\,3\,days is somewhat longer as the one determined previously 
(Van Winckel et al. 1995).  The vertical lines on the figure indicate the 
epochs of inferior and superior conjunction.

\begin{figure}
\begin{flushleft}
\mbox{\epsfxsize=3.2in\epsfysize=4in\epsfbox[117 345 470
710]{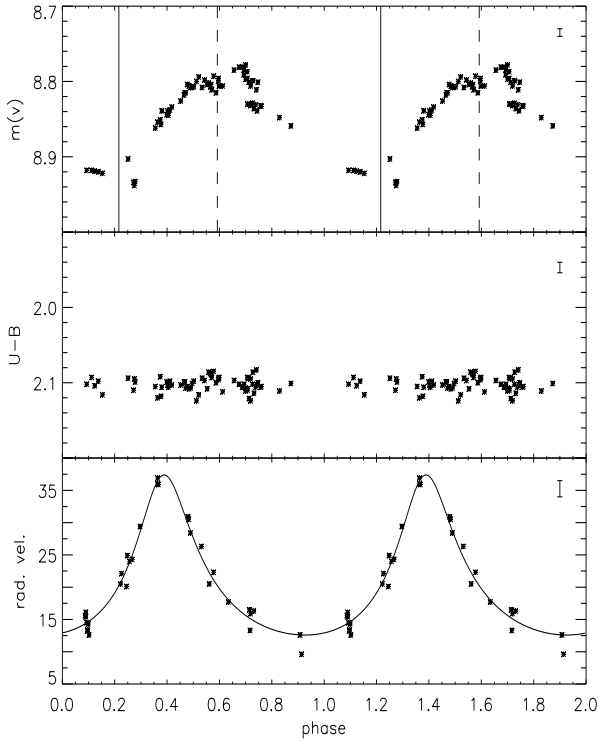}}
\caption[]{The phase diagram for the photometric and radial-velocity
variations of HD\,44179.  The phase of inferior(superior) conjunction is
marked by a full (dashed) vertical line. Phase 0 corresponds arbitrarily to JD
2448300. A typical error-bar for an individual measurement is shown in the 
upper-right corner of each panel.
}
\end{flushleft}
\end{figure}

It is apparent from Figure~2 that photometric variability occurs with the 
orbital
period.  As in the case of HR~4049, minimum brightness occurs at
inferior conjunction and maximum brightness at superior conjunction.  In the
case of HR~4049 the photometric variations are caused by variable obscuration
by the circumbinary disk during the orbital motion.  This interpretation was
confirmed by the colour variations, which are consistent with extinction.
However, in the case of HD~44179, {\it no color variations are observed, not in
[U-B], nor in any other color in the optical range}.  
If variable extinction along the line of sight
is responsible for the variability, then the grains causing 
it must be larger than in HR~4049.
On the other hand, many spectral features
attest the prominent presence of small grains in the Red Rectangle nebula.
We propose that the photometric variability of HD~44179 is not caused
by variable extinction, but by the
variability around the orbit of the scattering angle of the light that is
observed.

The two scattering clouds that are observed cannot be located on the orbital 
axis of the system, since then no orbital motion would be observed at all!  
It is much more likely that what we observe is light scattered from the
transition region between the optically thick disk and the optical nebula.
Roddier et al. (1995) found a smaller opening angle of the inner source 
(40$^{\circ}$) than for the nebula (70$^{\circ}$); this can be understood in 
our model: the scattering angle at the edge of the cone, as seen in projection,
is 90$^{\circ}$ and so probably too large for a significant flux in our 
direction; the light we observe, must be reflected by that part of the cone 
that is directed to us, i.e. where the scattering angle is smallest.  In the
following, we therefore assume that the inclination at which the orbital motion
is observed, is equal to half the opening angle of the cone, i.e. 35$^{\circ}$.

\begin{figure}
\begin{flushleft}
\mbox{\epsfxsize=3.4in\epsfysize=2.in\epsfbox{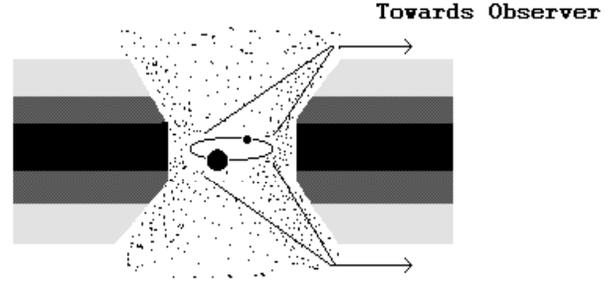}}
\caption[]{A geometric model for the inner part of the Red Rectangle.
Note that the thickness of the disk is two orders of magnitude
larger than the size of the binary system.  During the orbit, the scattering
angle toward the observer varies, being largest at inferior conjunction,
when minimum brightness is observed.}
\end{flushleft}
\end{figure}

Our model is schematically presented in Figure~3.  The orbital plane of the
binary is nearly edge-on, as is assumed commonly, since the star is hidden and
the nebula is remarkably symmetric. The variable scattering angle can be 
estimated from the size of the orbit
and the geometry of the nebula. Roddier et al. (1995) determined that the 
scattering clouds are located some 0.07" above and under the orbital plane.  
Assuming an absolute magnitude -4.0 for the star we then find from the observed
bolometric luminosity and reddening (Leinert \& Haas, 1989) a distance 
of 360\,pc, close to the value originally estimated by Cohen et al. (1975) 
on the basis of different assumptions.  This distance then implies that the
scattering occurs at a vertical distance of 25\,AU from the orbital plane;
with an opening angle of the nebula of 70$^{\circ}$, it follows that the 
distance of this region to the orbital axis is some 17.5\,AU.  From the 
orbital elements we derive that the radial distance of HD~44179 from the 
center of mass of the system amounts to 0.53\,A.U. at both inferior and 
superior conjunction, so that the scattering angle of the light we observe 
varies between 54.2$^{\circ}$ and 55.8$^{\circ}$.

It is customary to parametrize the scattering function S($\theta$) of an
astronomical source by a Heyney-Greenstein function of the form
$$
 S(\theta) = (1\,-\,g^{2})\,(1\,+\,g^{2}\,-2\,g\,cos\,\theta)^{-3/2}
$$
Isotropic scattering corresponds to \(g=0\) and $g$ approaches unity for strong
forward scattering.  In typical sources, $g$ ranges between 0.6 and 0.8.
For $g$ values of 0.6, 0.7, and 0.8, the ratio of forward scattered light to 
that scattered at an angle of 55$^{\circ}$ varies by factors 8.6, 21.1, and 
76.8, respectively; in the latter case, the brightness would be much lower than
is observed, so that it appears likely that $g$ falls in the range 0.6-0.7.
The variable scattering angle then induces photometric variations with an
amplitude between 0.067 and 0.076 mag.  The observed amplitude of some 0.12 mag
is slightly larger; nevertheless, the agreement with a model in which the 
scattering surface was assumed constant and no additional extinction variations
were taken into account, is encouraging.

\section{Evolutionary history of the system}

The mass function of the spectroscopic binary is 0.049\,$M_{\odot}$.  Assuming 
an `effective' inclination of 35$^{\circ}$, the mass of the unseen companion
can be derived for various masses of the primary.  For primary masses in
the range between 0.56 and 0.80 $M_{\odot}$, typical for post-AGB stars,
the mass of the secondary falls in the range betwen 0.77 and 0.91 $M_{\odot}$,
i.e. masses well below the initial mass of the primary.  It is then most
natural to assume that the secondary is a low-mass main-sequence star.

The present orbital parameters of the system are such that no AGB star with
the same luminosity can fit into the orbit.  On the other hand, if Roche-lobe 
overflow had occurred on the AGB, it is dubious whether the system could have 
survived as a relatively wide binary.  This problem is already encountered for 
HR\,4049, whose orbital period is 429 days.  It is even more severe in the 
case of HD\,44179, because its orbit is shorter, and moreover the initial mass 
of the star was probably larger than for HR\,4049.  Nevertheless, the carbon 
richness of the nebula does suggest that the star has gone through the thermal 
pulses which normally occur near the end of AGB evolution.

The present characteristics of the Red Rectangle therefore suggest that
filling of the Roche lobe was not necessary for mass transfer to occur.
Probably, then, mass loss started on the AGB before the Roche lobe
has been filled, altering substantially the evolution of HD~44179.
Apparently, this mass-loss process prevented the star from ever filling
its Roche lobe, even during the thermally pulsing AGB.  The luminosity may
have increased at a normal rate, but the radius would be lower than for single
AGB stars, i.e. the AGB evolution takes place on a track which is much bluer
than for single stars.

It is then not at all clear whether the present envelope mass is as low as
the 0.05 solar mass that is usually assumed at the start of a post-AGB
evolutionary track.  Indeed, in the scenario we propose, one may argue whether
these stars can be called {\it post}-AGB stars.  The present evolutionary 
timescale of HD~44179 could then be longer than for typical post-AGB stars.  
We note that a longer timescale is indeed more consistent with the huge extent 
and small outflow velocity of the nebula: from a coronograhic picture taken at 
the ESO NTT, the extent of the nebula is some 40" on both sides; the expansion 
velocity, deduced from the CO lines, amounts to some 6\,km/s (Jura et al. 
1995); hence, an age of more than 10\,000 years follows for the nebula, much 
longer than the typical post-AGB timescale.  A very short evolutionary 
timescale seems unlikely also in view of the fact that not less than four 
such objects brighter than nineth magnitude are known.

Similar problems with orbital sizes are well known for the barium stars, which
are binaries containing a white dwarf, which thus formerly was an AGB star.
The overabundances of the s-process elements in the barium stars are
interpreted as due to wind accretion from this AGB star.  Also barium stars
occur with orbital periods that are too short for normal AGB evolution.
It has been suggested that the progenitors of these close barium stars have
always remained detached, and that mass transfer occurred via wind accretion
(Boffin \& Jorissen 1988; Jorissen \& Boffin 1992; Theuns \& Jorissen 1993).

We suggest that HR~4049 and HD~44179 are progenitors of barium stars.
Indeed, they are binaries with similar periods, the primary of which is
presently finalizing its evolution before it becomes a planetary nebula and
then a white dwarf.  In our systems, the secondaries, that will later be
seen as barium stars, are still on the main sequence.  Important mass loss
is presently observed, and it is likely that a fraction of it is captured by
the companion.

Unfortunately, this suggestion cannot easily be checked directly.
The companion is much too faint to detect the s-process elements it accreted
after they had been produced during the thermal pulses of the primary.  Only
for one star of the group, the coolest member HD\,52961, could the s-process
elements Ba and Sr be detected in the spectrum of the primary: they follow
the iron abundance rather closely, because also these elements have been
absorbed by the dust and were not reaccreted.

Another peculiar characteristic which is shared by our systems and the
short-period barium stars, is the fact that the orbits are eccentric,
while one would expect that tidal interactions are very effective in
circularizing the orbits.  Here again, the substantial circumbinary disks
these objects develop may yield the answer.  Recent studies show that
binaries may acquire their eccentricity through tidal interactions with the
disks in which they are formed (Artymovicz et al. 1991).  It is then natural
to conjecture that the mass lost by HD~44179 and HR~4049, which appears to
accumulate preferentially in a disk, finally increases the eccentricity rather
than circularizing the orbit.  If this conjecture proves true, it would
strengthen the link with the barium stars, because it would imply that a
previous circumbinary disk must be invoked to explain the present
eccentricities of barium stars in close binary systems.

\acknowledgements{The authors thank Drs. Henny Lamers, Alain Jorissen, and
Ren\'e Oudmaijer fruitful discussions.  We thank the staff of
Geneva Observatory for their generous awarding of telescope time at the Swiss
Telescope at La Silla Observatory.  We also thank Dr. Hugo Schwarz for his
help for obtaining the coronographic image and Hans Plets for the fitting
of the velocity curve.  This work has been sponsored by the Belgian National
Fund for Scientific Research, under grant No. 2.0145.94.  LBFMW gratefully
acknowledges support from the Royal Netherlands Academy of Arts and Sciences.
EJB was supported by grant No. 782-371-040 by ASTRON, which receives funds
from the Netherlands Organisation for the Advancement of Pure Research (NWO).}

\end{document}